# Alerts in High-resolution TEM characterization of perovskite material

**Yu-Hao Deng[1*]**

[1] Academy for Advanced Interdisciplinary Studies, Peking University, Beijing, China

* Correspondence: yuhaodeng@pku.edu.cn

**Abstract:** High-resolution TEM (HRTEM) is a powerful tool for structure characterization. However, methylammonium lead iodide ($MAPbI_3$) perovskite is highly sensitive to electron beams and easily decompose into lead iodide ($PbI_2$). Universal misidentifications that $PbI_2$ is incorrectly labeled as perovskite are widely exist in HRTEM characterization, which would negatively affect the development of perovskite research field. Here misidentifications in $MAPbI_3$ perovskite calibration are summarized, classified and corrected based on corresponding electron diffraction (ED) simulations. Corresponding crystallographic parameters of intrinsic tetragonal $MAPbI_3$ and the confusable hexagonal $PbI_2$ are also presented clearly. Finally, the method of proper phase identification and some ways to control the radiation damage in HRTEM are provided. This work paves the way to avoid misleadings in HRTEM characterization of perovskite and other electron beam-sensitive materials in the future.

## Introduction

High-resolution transmission electron microscopy (HRTEM) is a very powerful characterization tool and has been extensively and successfully used for analyzing crystal structures on an atomic resolution scale [1-4]. Recently, halide perovskites have achieved substantial success in various optoelectronic devices owing to their solution-based growth method and remarkable physical properties [5-9]. However, the best-known methylammonium lead iodide ($MAPbI_3$) perovskite is very sensitive to electron beam irradiation. Figure 1 shows the $MAPbI_3$ degradation process under electron beam irradiation. Tetragonal perovskite decomposes into hexagonal lead iodide ($PbI_2$) by the escape of methylamine and hydrogen iodide molecules.

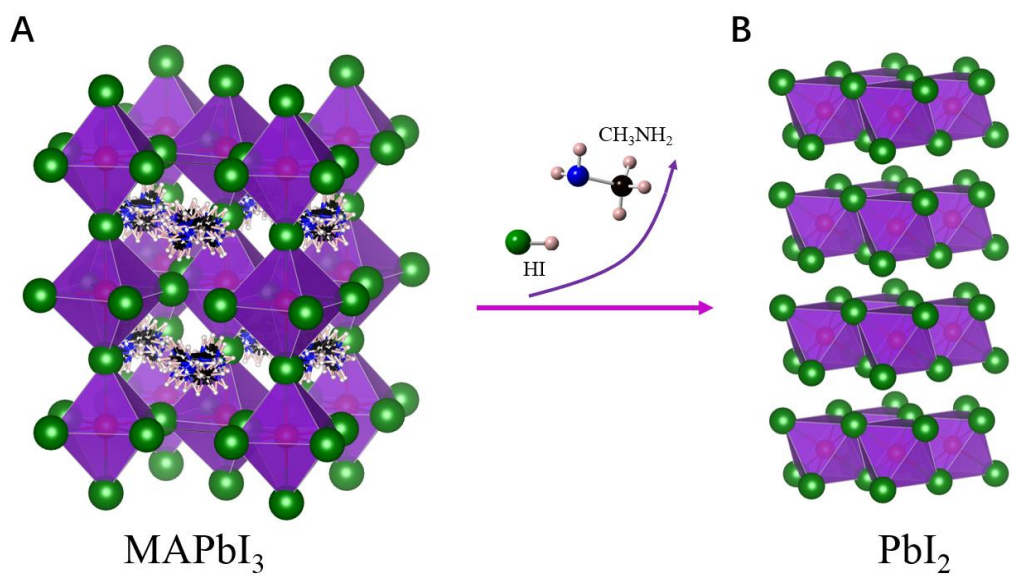

**Fig. 1. MAPbI₃ degradation under electron beam irradiation.** Tetragonal perovskite (**A**) decomposes into hexagonal lead iodide (**B**), followed by the escape of methylamine and hydrogen iodide molecules. Here, colors represent the following: green, iodine; red, lead; black, carbon; blue, nitrogen; pink, hydrogen.

Comparing a group of interplanar spacing and angles to identify phase is easy to ignore the missing crystal planes and lead to mistaken identification results, such as lead iodide would be generally labeled as perovskite. In a proper phase identification, HRTEM images cannot alone be used for phase identification, but always has to be analysed together with the relevant diffractograms, simulated ED, nanodiffractions, or XRD specimen data etc [1]. As a rough estimation, MAPbI₃ begins to decompose into PbI₂ under 150 eÅ⁻² total dose irradiation [10, 11]. However, the value of the electron dose in normal HRTEM is around 800-2000 eÅ⁻² s⁻¹, which is much higher than the critical dose of MAPbI₃ perovskite. Meanwhile, some interplanar spacing and angles in the decomposition product (PbI₂) are very similar with MAPbI₃, so PbI₂ is easy to be misidentified as perovskite. A typical feature of this misidentification is the missing of crystal planes in the results. Therefore, the absence of crystal planes indicates that the material would be no longer MAPbI₃ perovskite, but other phases and structures. Particularly, the missing crystal planes have been observed in low-dose imaging. Thus far, Zhu, etl. have got the missing (1$\bar{1}$2), (112) planes with 0.44 nm along the [201] zone axis by Cryo-TEM under condition of low electron dose is about 3 e Å⁻² [12]. Similarly, Song, etl. have got the missing (1$\bar{1}$0), (002) planes with 0.63 nm along [110] zone axis at total doses of 1.5 eÅ⁻² [13]. Moreover, the missing crystal planes have also been observed in other characterization techniques based on the Bragg's law, such as SEAD and XRD [14-17]. Finally, the low-order crystal plane (d(112)= 4.4 Å) has longer interplanar spacing than the high-order crystal plane (d(224)= 2.2 Å). If the low-order crystal plane exists and the diffraction intensity is enough, then the low-order crystal plane should be observed. So, the missing of low-order crystal planes should not result from the poor accuracy in the data interpretation. Although the mistakes are being taken seriously [10, 11, 36], a most complete summary and correction is still highly urgent.

## Results and discussion

Here misidentifications in MAPbI$_3$ perovskite are summarized, classified and corrected. Figure 2 shows simulated ED patterns of MAPbI$_3$ and PbI$_2$ along different axis zones. Figure 2A is the ED pattern of MAPbI$_3$ along [110] zone axis. ($\bar{1}$10), (002) crystal planes are existing in the intrinsic tetragonal perovskite, but they are missing in HRTEM characterizations in some papers [18-21]. Figure 2B shows the simulated ED patterns of PbI$_2$ along [4$\bar{4}$1] zone axis. (014), ($\bar{1}$04) crystal planes of PbI$_2$ have the confusable interplanar spacing and angle with ($\bar{2}$20), (004) crystal planes of MAPbI$_3$. Actually MAPbI$_3$ would have been damaged into PbI$_2$ phase by electron beams and ignoring the absence of crystal planes results in the mischaracterization. Similarly, Figure 2C-H are ED patterns of MAPbI$_3$ along [101] and PbI$_2$ along [8 10 1], MAPbI$_3$ along [$\bar{2}$01] and PbI$_2$ along [8$\bar{8}$1], MAPbI$_3$ along [$\bar{1}$20] and PbI$_2$ along [$\bar{4}$11] respectively. The missing crystal planes in HRTEM characterizations have been circled in red in Figure 2.

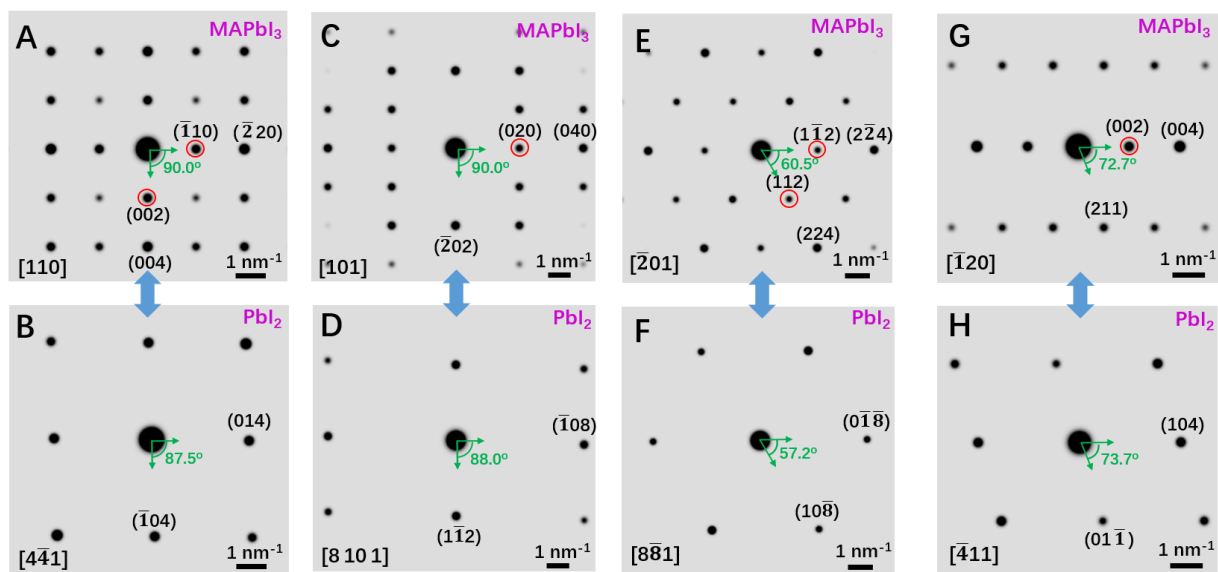

**Fig. 2. Simulated electron diffraction (ED) patterns of tetragonal MAPbI$_3$ and hexagonal PbI$_2$. (A)** MAPbI$_3$ along [110] axis zone. **(B)** PbI$_2$ along [4$\bar{4}$1] zone axis. **(C)** MAPbI$_3$ along [101] axis zone. **(D)** PbI$_2$ along [8 10 1] zone axis. **(E)** MAPbI$_3$ along [$\bar{2}$01] axis zone. **(F)** PbI$_2$ along [8$\bar{8}$1] zone axis. **(G)** MAPbI$_3$ along [$\bar{1}$20] zone axis. **(H)** PbI$_2$ along [$\bar{4}$11] zone axis. Crystal planes marked in red circle are missing in published articles [18-38].

The misidentifications exist in the following fields: phase identification and structure determination [22], grain, nanowire and microwire orientation [20, 23-26], morphology analysis and shape control of nanocrystals [27], growth direction of perovskite materials [28], degradation process and kinetics of perovskite [29], phase transition research in perovskite [30] and ion migration characterization in photoelectric devices [31]. Not only in single component of MAPbI$_3$ material, the misidentifications also occur in the field of heterostructure materials. Such as lattice matching and kinetic study in epitaxial growth of perovskite films on 2D material [32], PbS quantum dots in perovskite material [18, 19] and lattice anchoring stabilizes perovskite research [20]. Some of those articles are about the performances of optoelectronic devices. HRTEM is an auxiliary tool to identify the phase of perovskite materials and misinterpretations would not influence their device performances and conclusions. However, in some cases, such as the growth direction, orientation, and lattice matching of perovskite materials. Misinterpretations in HRTEM would result in wrong conclusions. To make the comparisons and corrections clearer, Table 1 shows the detailed parameters of the intrinsic MAPbI$_3$ and confusable PbI$_2$ along different zone axis, missing crystal planes have been marked in red. Even remarkably, the [110] and [001] zone axis are equivalent in tetragonal MAPbI$_3$ perovskite.

Table 1. Detailed crystallographic parameters of MAPbI$_3$ and PbI$_2$ along different zone axis corresponding to Fig. 2, the missing low-order crystal planes of MAPbI$_3$ have been marked in red. The close angles and interplanar spacing make them easily confused.

| Material and zone axis | Characteristic crystal planes | Interplanar spacing | Interplanar Angle | References |
|---|---|---|---|---|
| MAPbI$_3$ [110] | ($\bar{1}$10), ($\bar{2}$20) (002), (004) | d($\bar{1}$10)= 6.2Å. d($\bar{2}$20)= 3.1Å. d(002)= 6.3Å. d(004) =3.2Å. | <($\bar{1}$10), (002)> = <($\bar{2}$20), (004)> =90.0º | 21-32 |
| PbI$_2$ [4$\bar{4}$1] | (014) ($\bar{1}$04) | d(014)= 3.2Å. d($\bar{1}$04)= 3.2Å. | <(014), ($\bar{1}$04)> =87.5º | ________ |
| MAPbI$_3$ [101] | (020), (040) ($\bar{2}$02) | d(020)= 4.4Å. d(040)= 2.2Å. d($\bar{2}$02)= 3.6Å. | <(020), ($\bar{2}$02)> = <(040), ($\bar{2}$02)> =90.0º | 33, 34 |
| PbI$_2$ [8 10 1] | ($\bar{1}$08) <br> (1$\bar{1}$2) | d($\bar{1}$08)= 2.2Å. <br> d(1$\bar{1}$2)= 3.7Å. | <($\bar{1}$08), (1$\bar{1}$2)> =88.0º | _______ |
| MAPbI$_3$ [$\bar{2}$01] | (1$\bar{1}$2), (2$\bar{2}$4) <br> (112), (224) | d(1$\bar{1}$2)= 4.4Å. <br> d(2$\bar{2}$4)= 2.2Å. <br> d(112)= 4.4Å. <br> d(224)= 2.2Å. | <(1$\bar{1}$2), (112)> = <(2$\bar{2}$4), (224)> =60.5º | 18-20, 24 |
| PbI$_2$ [8$\bar{8}$1] | (0$\bar{1}$ $\bar{8}$) <br> (10$\bar{8}$) | d(0$\bar{1}$ $\bar{8}$)= 2.2Å. <br> d(10$\bar{8}$)= 2.2Å. | <(0$\bar{1}$ $\bar{8}$), (10$\bar{8}$) > =57.2º | _______ |
| MAPbI$_3$ [$\bar{1}$20] | (002), (004) <br> (211) | d(002)= 6.3Å. <br> d(004) =3.2Å. <br> d(211)= 3.8Å. | <(002), (211)> = <(004), (211)> =72.7º | 35-38 |
| PbI$_2$ [$\bar{4}$11] | (104) <br> (01$\bar{1}$) | d(104)= 3.2Å. <br> d(01$\bar{1}$)= 3.9Å. | <(104), (01$\bar{1}$) > =73.7º | _______ |

The MAPbI$_3$ perovskite is extremely beam-sensitive with a critical dose around a few hundred electrons per square angstrom. For most measurements, and in particular HRTEM, keeping the dose below the critical value is extremely challenging [39-41]. Herein, there are some ways to control the radiation damage in HRTEM, and they might be helpful in obtaining the intrinsic structure of perovskite materials. Low-dose imaging techniques are effective approaches for electron beam-sensitive materials [42]. Zhang etl. developed a program to achieve a direct, one-step alignment of the zone axis and got the HRTEM images of electron beam-sensitive materials under 6-12 e Å$^{-2}$ [43]. Carlino reported the in-line holography in TEM for the study of radiation-sensitive materials providing high-contrast holograms under 1-2 eÅ$^{-2}$s$^{-1}$ [44]. The intrinsic structure of MAPbI$_3$ has been captured by the Direct-detection electron-counting (DDEC) cameras under 3 e Å$^{-2}$ [12, 45]. Besides, cool the specimen is also a feasible way and HRTEM images of MAPbI$_3$ has been captured by the cryo-TEM [11, 12]. Finally, encapsulation of specimen with a thin, stable and continuous layer would help to reduce radiation damage [41]. In any case, the total electron dose should be below the critical dose of MAPbI$_3$, otherwise substantial damage occurs during TEM characterizations.

**Conclusions**

The MAPbI$_3$ perovskite is very sensitive to electron beam irradiation and easily decompose into PbI$_2$. Here summary and analysis of misidentifications in HRTEM characterization of MAPbI$_3$ material are very helpful for researchers to avoid mistakes in perovskite research field. Lessons learned from the mistakes alert us it is unreliable to calibrate material phase only by measuring interplanar spacing and angles. We also need to refer the diffractograms, simulated ED, nanodiffractions or XRD data to ensure the crystal planes are complete. Particularly, the method of proper phase identification and some ways to control the radiation damage in HRTEM are provided in this paper. This work provides a sober mind for further characterization in organic-inorganic hybrid perovskite and other electron beam-sensitive materials.

**Methods**

Corresponding crystal structures cif files were downloaded from Crystallography Open Database (COD) website. COD IDs of MAPbI$_3$ and PbI$_2$ are 4124388 and 9009141 respectively [46, 47]. MAPbI$_3$ is I4/mcm space group with tetragonal structure, cell parameters: a=b=8.839Å, c=12.695Å; α=β=γ=90°. PbI$_2$ is P-3m1 space group with hexagonal structure, cell parameters: a=b=4.555Å, c=20.937Å; α=β=90°, γ=120°. The Electron diffraction (ED) simulations of MAPbI$_3$ and PbI$_2$ were obtained using CrystalMaker Software. The interplanar spacing and interplanar angle can be calculated from the cell parameters. During the process of phase identification, I also tried other polytypes of PbI$_2$, but the results could not match well [48].

**Data availability:** All data are available from the corresponding author(s) upon reasonable request.

**Conflict of interest:** The authors declare no competing financial interest.

**Note added:** This work has been published in ***Condensed Matter***: Deng, Y.-H. Common Phase and Structure Misidentifications in High-Resolution TEM Characterization of Perovskite Materials. *Condens. Matter* 2021, 6, 1. https://doi.org/10.3390/condmat6010001